# Optical detection of the sliding ferroelectric switching in hBN with a WSe$_2$ monolayer

Sébastien Roux[1,*], Jules Fraunié[1], Kenji Watanabe[2], Takashi Taniguchi[3], Benjamin Lassagne[1], Cedric Robert[1,*]

[1]Université de Toulouse, INSA-CNRS-UPS, LPCNO, 135 Av. Rangueil, 31077 Toulouse, France.
[2]Research Center for Electronic and Optical Materials, National Institute for Materials Science, 1-1 Namiki, Tsukuba 305-0044, Japan
[3]Research Center for Materials Nanoarchitectonics, National Institute for Materials Science,  1-1 Namiki, Tsukuba 305-0044, Japan

Abstract: *When two BN layers are stacked in parallel in an AB or BA arrangement, a spontaneous out-of-plane electric polarization arises due to charge transfer in the out-of-plane B-N bonds. The ferroelectric switching from AB to BA (or BA to AB) can be achieved with a relatively small out-of-plane electric field through the in-plane sliding of one atomic layer over the other. However, the optical detection of such ferroelectric switching in hBN has not yet been demonstrated. In this study, we utilize an adjacent WSe$_2$ monolayer to detect the ferroelectric switching in BN. This dynamic coupling between a 2D ferroelectric and a 2D semiconductor allows for the fundamental investigation of the ferroelectric material using a non-destructive, local optical probe, offering promising applications for compact and non-volatile memory devices.*

Ferroelectric (FE) materials exhibiting spontaneous electric polarization have been considered as promising candidates for a new generation of electronic devices, including non-volatile memories, transistors and neuromorphic components [1]. For the past decades, most of the studies have focused on FE materials with three-dimensional (3D) crystal structures. Nevertheless, their use in practical devices requires epitaxial growth of thin films. Unfortunately, the reduction of thickness generally goes along with an increase of the depolarizing field. Moreover, the integration of these traditional FE materials is limited by the lattice-matching constraint with their grown substrate. The recent emergence of two-dimensional (2D) FE materials made of van der Waals (vdW) crystals can potentially provide technological solutions as these layers can be easily interfaced with any other material. CuInP$_2$S$_6$ and In$_2$Se$_3$ are probably the two most studied 2D FE materials for the past few years [2,3] with demonstration of electronic device prototypes [4–12].

In 2017, it was theoretically predicted that out-of-plane electric polarization can also emerge in parallel alignment of bilayers of some van der Waals crystals that are normally not ferroelectric [13]. The experimental demonstrations came in 2021, when FE was observed in marginally twisted layers of hexagonal boron nitride (hBN) [14–16]. Shortly after, similar observations were made with parallelly aligned transition metal dichalcogenide (TMD) bilayers [17,18] and even with graphene [19]. Intriguingly, the switching mechanism between upwards and downwards polarization states results from an in-plane interlayer sliding of one layer with respect to the other. These sliding FE structures have several advantages. Their unique switching mechanism gives rise to low switching barriers (few meV/formula unit) [13,14,20,21] that are promising for high-speed data-writing at low energy cost. The mechanism is not limited to bilayers but has been shown to be cumulative in multilayers [16,22] which is interesting for multistate memories and neuromorphic applications. Sliding FE is compatible with room temperature operation and well above [23]. Finally, the mechanism has been demonstrated in a wide range of materials from insulators (hBN), semiconductors (TMD) and metals (graphene) that allows for exploring disruptive multifunctional device concepts with electrical and/or optical control.

The combination of FE and optical properties is interesting for photovoltaic sensors, optically addressable memories and optoelectronic synapses[24,25]. A potential approach would be to combine ferroelectric (FE) and semiconducting (SC) properties, as seen in 3R TMD bilayers [26,27]. In this work, we explore an alternative strategy by combining insulating FE hBN with a direct bandgap SC $WSe_2$ monolayer (ML). hBN is totally air insensitive and is one of the most used 2D materials in vdW heterostructures mainly for its excellent encapsulation properties[28,29]. The use of FE insulators like hBN offers distinct advantages, particularly as a dielectric gate material in Ferroelectric Field Effect Transistors (Fe-FETs)[30]. Notably, insulating FE hBN exhibits the highest polarization among 2D sliding FE materials [17]. Moreover, while the polarization of sliding FE materials has been shown to accumulate with the number of interfaces [22], it ultimately saturates due to bandgap closure [31] With a large bandgap of 6 eV, rhombohedral BN multilayers are expected to have a significantly higher polarization limit compared to rhombohedral TMDs. Finally, sliding hBN has recently been shown to have exceptionally long FE operation times[32]. Despite these compelling advantages, the optical detection of the sliding phase transition in hBN has not yet been demonstrated.

Only very few studies have dealt with FE hBN and TMD ML. Zhao *et al.* theoretically predicted that a hBN interface with a nearly parallel alignment can create a superlattice potential in a TMD ML[33]. D. Kim et al., then claimed that this superlattice potential impedes exciton diffusion in a $MoSe_2$ ML[34]. In our previous work, we showed that parallel alignment in a self-folded hBN flake results in AB and BA ferroelectric domains, sketched in Fig. 1 (a), with upwards and downwards polarization that imprint n and p doped domains in a remote $WSe_2$ ML [35]. However, in these previous studies, the TMD was not electrically contacted, and the FE switching of hBN was not actively controlled. As a result, the extent to which the FE domains modulate the doping density in the TMD monolayer was only partially explored.

In this letter, using a graphite-FE hBN-$WSe_2$ device, we show that the $WSe_2$ ML can be used as an optical probe of the domain switching at the hBN FE interface. We apply an out-of-plane electric field to switch the FE domains that can be monitored by the photoluminescence (PL) intensity of charged excitonic complexes in $WSe_2$ ML. The TMD luminescence is then used to probe the hysteresis cycles of the FE hBN interface as well as the effect of the FE phase switching on the capacitive interaction between the $WSe_2$ ML and the graphite gate. Our optical probing approach has several advantages as compared to the very few existing studies on switching in sliding FE materials. Our spatial resolution is only limited by optical diffraction which is better than with the full electrical probing using the conduction in a graphene sensor layer [14,32]. It also presents some advantages in terms of simplicity of device fabrication as compared to the device required in operando electronic microscopy experiments that studied switching [15,21]. Moreover, the optical probing could be used in the future to study fast switching dynamics and domain walls translation.

The sample is sketched in Figure 1(b). The FE interface is created by tearing and stacking two pieces of a 6-nm thick hBN flake using a dry transfer method with a PDMS stamp, as detailed in the Supporting Information[36]. After checking the FE properties by Kelvin Force Microscopy (KFM) (see Supporting Information), we deposit a $WSe_2$ ML and a thin hBN top layer. Using a pick-up method with a PVC film [37], this stack is then transferred on a thin graphite flake acting as a back gate and covered with a 110 nm thick hBN flake. The $WSe_2$ ML and the back gate are contacted by Ti/Au electrodes pre-patterned on a $SiO_2$/Si substrate (More details on the fabrication and on the KFM characterization are presented in the Supporting Information). Figure 1(c) presents the KFM image of the final structure before applying any out-of-plane electric field. The image is zoomed on a region with large domains of several µm (highlighted by yellow lines) where we focus our study. The KFM and optical images of the full sample are shown in the Supporting Information. We identify an alternance of dark blue and light blue domains mainly with triangular shapes and with size ranging from 100 nm to several µm. This corresponds to FE domains resulting from local AB and BA parallel alignment at the hBN interface. The line scan of the red solid line is presented in Figure 1(d) and shows a surface potential contrast of around 115 mV between AB and BA domains, in agreement with the literature [16,34,35].

Figure 1 (e) shows typical PL spectra of the $WSe_2$ ML emission taken at 4 K and measured above the AB domain, in dark blue, and above the BA domain, in light blue. The reference energy is fixed at the position of the neutral exciton $X^0$. As shown in ref. [35], the ferroelectric domains induce a modulation of the doping on the $WSe_2$ monolayer. Here, the doping modulation is smaller than that in Ref. [35] which was conducted at 55 K and achieved both p-type and n-type doping. Still, the electron doping density increases when $WSe_2$ is stacked on top of a BA domain compared to when stacked on top of an AB domain. This is clearly observed on the spectra by the drastic increase in the PL intensity of the negatively charged excitons (two electrons bound with a single hole), namely the triplet negative trion ($X_T^-$) and the singlet negative trion ($X_S^-$). The ratio $X_S^- / X^0$ can then be used to monitor the negative doping level of the layer and thus to locally detect the FE domain polarization. This is shown by the PL mapping of $X_S^- / X^0$ in Fig. 1 (f) (experimental details in the Supporting Information), where one can recognize the larger domains that appear in the KFM image in (c).

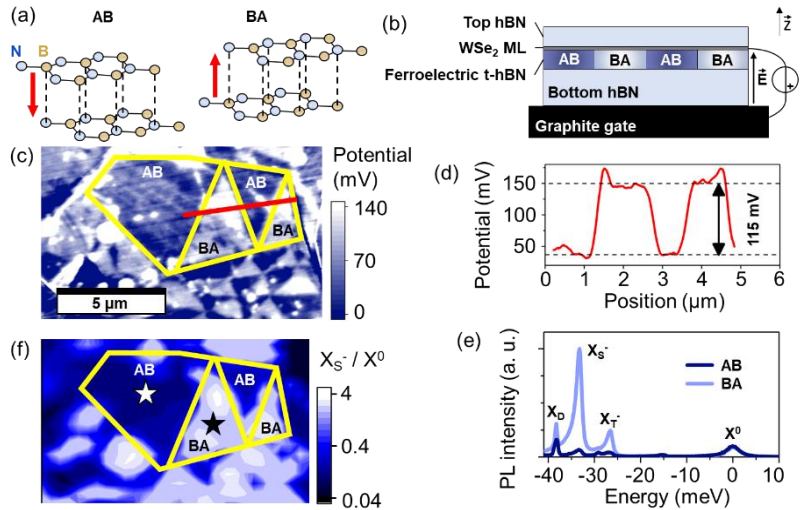

*Fig. 1. (a) Scheme of the two stable parallel alignment of BN layers : AB and BA. The red arrows represent the electric polarizations that emerge from the charge transfer between the two layers. (b) Sketch of the sample. (c) KFM image made on the top surface of the whole structure. (d) Line scan of the KFM potential measured along the red line in (c). (e) PL spectra of the $WSe_2$ ML luminescence measured under a 10 µW, 1.96 eV excitation at T=4 K on a BA domain and on an AB domain before any bias is applied between the two electrodes. (f) Mapping of the ratio between the integrated PL intensity of the negative trion $X_S^-$ divided by the one of the neutral exciton $X^0$. The white and black stars indicate the positions of the spectra in (e). The larger domains are surrounded by solid yellow lines in (b) and (e). AB domains appears in dark blue, while BA domains in light blue.*

Now that the sample is primarily characterized, the effect of the bias is investigated. Note that our structure is a single gate device. Thus, applying a bias both creates an out-of-plane electric field but also electrostatically tunes the free carrier density in the $WSe_2$ ML. In this manuscript, we consider the bias applied to the graphite backgate and the z-axis pointing upwards, such that a positive bias is associated with a positive electric field pointing upwards (see Fig. 1(b)) and a negative doping of the $WSe_2$ ML. Figure 2(a) and (b) show color plots of the PL intensity as a function of the applied bias in the small carrier density regime, between -0.5 V and +0.5 V, corresponding to a maximum carrier density of around $8*10^{10}$ cm$^{-2}$ (details in the Supporting Information) and electric field of about 0.004 V.nm$^{-1}$. Additional data are shown in Supporting Information. We recognized the neutral exciton peak $X^0$ and the two negative trions $X_S^-$ and $X_T^-$. The color-plot shown in Figure 2(a) was recorded after application of a large bias of +13 V corresponding to an electric field of +0.11 V.nm$^{-1}$ pointing upwards while the color-plot depicted in Figure 2(b) was recorded after applying a large bias of –13 V corresponding to an electric field of –0.11 V.nm$^{-1}$ pointing downwards. These fields are above the switching field of the hBN FE interface that is expected to be about 0.1 V.nm$^{-1}$ [14]. Both color plots were recorded at the same spot on the sample (white star in Fig. 1(f)).

It can be observed that the two color-plots are vertically shifted by about 200 mV, as shown in Figure 2(a) and 2(b) by the black solid horizontal lines representing the voltage at which the $X_S^-$ exciton luminescence appears. This is consistent with a FE switching from a BA to a AB phase, since the 200 mV shift is comparable with the surface potential contrast measured in KFM [16,34,35] (see Fig. 1 (c)).

Figure 2(c) shows the ratio $X_S^-$ / $X^0$ as a function of voltage after applying +0.11 V.nm$^{-1}$ and – 0.11 V.nm$^{-1}$, extracted from Figure 2(a) and 2(b). The same experiments were performed after applying smaller electric fields of ± 0.06 V.nm$^{-1}$, *i.e.* with out-of-plane fields smaller than the switching field of the ferroelectric interface (Figure 2(d)). This shows that the 200-mV shift observed between Figure 2(a) and 2(b) only occurs when the electric fields are large enough, strengthening the attribution of this shift to a domain switching. We show in the Supporting Information that there is no such shift when the same measurements are performed on a similar device without FE interface.

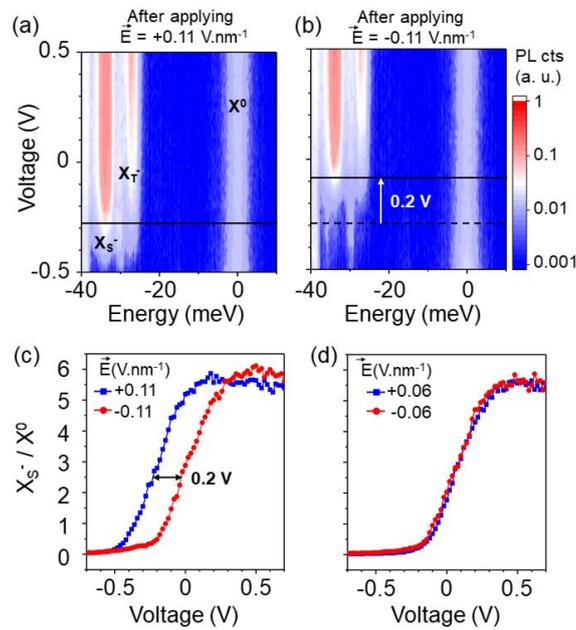

*Fig. 2 (a-b) PL spectra as a function of gate voltage after application of an out-of-plane electric field of +0.11 V.nm$^{-1}$ (a) and -0.11 V.nm$^{-1}$ (b), measured at 1 µW. (c-d) $X_S^-$ / $X^0$ as a function of the gate voltage and after applying ±0.11 V.nm$^{-1}$ (c) and after applying ±0.06 V.nm$^{-1}$ (d).*

Note that Figure 2(c) does not represent a hysteresis cycle, but only a 0.2 V lateral shift of the two curves. Our interpretation of the shift is the following: the FE polarization modulates the capacitive coupling between the graphite back gate and the WSe$_2$ ML, shifting the potential difference by approximately +100 mV and -100 mV for the AB and BA domains, respectively. Using a capacitor plate model, we estimate a corresponding modulation of electron carrier density on the order of $3 \times 10^{10}$ cm$^{-2}$ which is probably in the range of the intrinsic doping density in the WSe$_2$ ML. This is qualitatively consistent with the small variation observed in the ratio of negative trions to neutral exciton intensities. This figure also shows that above 0.4 V and below -0.4 V, the effect of the FE domain on the doping density inside the WSe$_2$ ML becomes negligible as compared to the electrostatic doping induced by the gate. Thus, the optical probing of the FE polarization must be performed close to the WSe$_2$ ML neutrality.

Figure 3(a) shows a mapping of the $X_S^-$ / $X^0$ PL ratio measured at 0V after application of a +0.14 V.nm$^{-1}$ electric field (pointing upwards). In the region highlighted by the yellow solid lines we observe an alternance of two white domains (BA stacking in hBN with upwards polarization) and two dark blue domains (AB stacking in hBN with downwards polarization). Figure 3(b) shows the same PL map, but this time measured after application of an electric field of –0.14 V.nm$^{-1}$ (pointing downwards). Here, the area of the BA domains has been divided by more than 4, so that most of the area is now an AB domain, including the point marked with the white star. The two maps demonstrate the optical detection of the FE switching in hBN by the WSe$_2$ ML luminescence.

Figure 3(c) shows the ratio $X_S^-$ / $X^0$ measured at 0 V (*i.e.* under a zero out-of-plane electric field) at the fixed position of the white star in Figure 3(a) and 3(b), as a function of the out-of-plane electric field applied prior to the measurement of each data point. The higher doping level (high $X_S^-$ / $X^0$ level) corresponds to a BA phase, while a low negative doping level (low $X_S^-$ / $X^0$) corresponds to an AB phase. Four square hysteresis cycles are clearly observed. Following the black and red arrows on Figure 3(c) we observe an extremely sharp transition from AB to BA phase at +0.065±0.001 V.nm$^{-1}$ (black arrow, see zoom on this transition in Fig3(d)) and from BA to AB at –0.098±0.002 V.nm$^{-1}$ (red arrow). Remarkably, these hysteresis cycles are extremely reproducible.

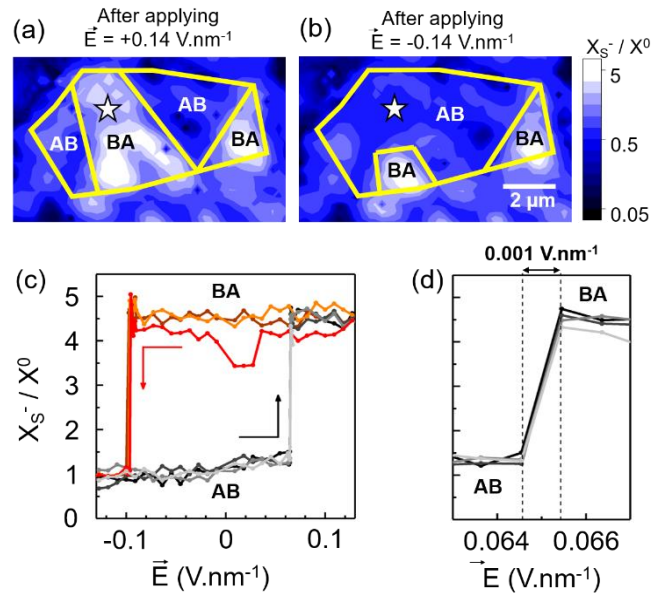

*Fig. 3. PL mapping of $X_S^-$ / $X^0$ measured at 1 µW after application of an out-of-plane electric field of +0.14 V.nm$^{-1}$ (a) and -0.14 V.nm$^{-1}$ (b). The domain walls are marked by the solid yellow lines in (a) and (b). (c) $X_S^-$ / $X^0$ measured at 0V and 1 µW as a function of the electric field applied before the measurement. (d) Zoom of the figure (c) around +0.065 V.nm$^{-1}$. This measurement was recorded on the white star that appears in (a) and (b).*

Several observations can be made from our results. The first obvious one is that the domain switching is not complete in Figure 3(a) and 3(b). After, the positive (negative) electric field, small AB (BA) domains persist. This agrees with previous studies on switching in sliding FE materials [15,21] that demonstrate that the corners of triangular shape domains can act as pinning centers and that only large domains can really fully switch. The second obvious observation is that the hysteresis cycle shown in Figure 3(c) is horizontally shifted by 16.5 mV.nm$^{-1}$. Such a shift is generally considered as a signature of imprint in FE materials [38]. We note that similar imprints have been observed in ref. [14]. The first possible reason for imprint is the asymmetry of the two electrodes. When two electrodes with different work functions $\varphi_1$ and $\varphi_2$ sandwich an insulator of thickness $d$, it creates a built-in electric field $F_i = (\varphi_2 - \varphi_1)/ed$ where $e$ is the electronic charge [39]. In Ref [14], the two electrodes are made of Au and graphene and the thickness is 9.6 nm which is consistent with an observed imprint of 50 mV.nm$^{-1}$. In our work, the two electrodes are made of graphite ($\varphi = 4.7$ eV [40]) and WSe$_2$ ($\varphi = 4.5$ eV [41]) and the hBN thickness is 122 nm. This should induce an imprint of the order of 1 mV.nm$^{-1}$ which is much smaller than what we observe in Figure 3(c). This indicates that there is another source of imprint such as pinning centers, dipolar defects in the hBN or fixed dipoles at the interface between graphite and hBN and/or WSe$_2$ and hBN [38,42,43]. Further work will be required to elucidate the exact origin of the imprint.

In summary, we have demonstrated that a remote WSe$_2$ ML can be used as an optical probe of polarization switching in a sliding FE hBN interface. We have observed the switching of a large domain and measured a reproducible square hysteresis cycle by measuring the PL intensity ratio of negative trion with respect to the neutral exciton. A significant imprint is observed in the hysteresis cycle that could be due to pinning centers or fixed electric dipoles at the interface between hBN and graphite or WSe$_2$. Future perspectives include the fabrication of double gated devices to control independently the out-of-plane electric field and the electrostatic doping in the WSe$_2$ ML. Such device could be used to optically study switching dynamics and measure domain walls speed as electric field could be applied while keeping the WSe$_2$ ML close to neutrality. Reducing the distance between the WSe$_2$ ML and the FE interface would also be interesting to increase the coupling between electric charges and dipoles in the SC with the polarization in the FE. Finally, this TMD/FE hBN vdW heterostructure is attractive to study the potential optical control of FE hBN polarization with the photogenerated carriers in the TMD layer.

Supporting Information:
Description of the sample fabrication, details on the experimental setup. Calculation of the carrier density as a function of the gate voltage and additional data showing PL color maps in a large doping range and in a sample without FE domains.

Corresponding authors:
*Sébastien Roux: sroux@insa-toulouse.fr
*Cedric Robert: cerobert@insa-toulouse.fr

Acknowledgements:
We thank Jean-François Dayen and Manuel Bibes for fruitful discussions and Louis Pacheco and Emmanuel Lepleux from CSI Instruments-Scientec for their help with KFM measurements. This work was supported by Agence Nationale de la Recherche funding under the program ESR/EquipEx+ (Grant No. ANR-21-ESRE-0025); ANR ATOEMS and ANR IXTASE, through Grant No. NanoX ANR-17-EURE-0009 in the framework of the "Programme des Investissements d'Avenir"; and by the Institute for Quantum Technologies in Occitanie through Project 2D-QSens. K.W. and T.T. acknowledge support from the JSPS KAKENHI (Grant Numbers 21H05233 and 23H02052) and World Premier International Research Center Initiative (WPI), MEXT, Japan.

# Supporting information

# Optical detection of the sliding ferroelectric switching in hBN with a WSe$_2$ monolayer

Sébastien Roux[1], Jules Frauníé[1], Kenji Watanabe[2], Takashi Taniguchi[3], Benjamin Lassagne[1], Cedric Robert[1]

[1]Université de Toulouse, INSA-CNRS-UPS, LPCNO, 135 Av. Rangueil, 31077 Toulouse, France.
[2]Research Center for Electronic and Optical Materials, National Institute for Materials Science, 1-1 Namiki, Tsukuba 305-0044, Japan
[3]Research Center for Materials Nanoarchitectonics, National Institute for Materials Science, 1-1 Namiki, Tsukuba 305-0044, Japan

**Sample preparation:**

The FE hBN stack, depicted in Figure S1 (a) was fabricated using a tear and stack method adapted for an all-dry process with a PDMS stamp [1]. First, a large, thin flake is optically identified on the PDMS stamp. The stamp is then partially brought into contact with a SiO$_2$/Si substrate so that only a part of the hBN flake touches the substrate. The PDMS is subsequently removed from the substrate. Eventually, one part of the hBN flake stays on the Si substrate while the second part stays on the PDMS. The two parts of the hBN are then stacked using the traditional transfer method [1]. Note that the efficiency of this tear and stack method increases with the size and the thinness of the flake. Here we use a large flake of about 100*100 µm² with a thickness of 6 nm. No twist angle is introduced during this process to maximize the size of the FE domains. The hBN-hBN stack is still marginally twisted due to the strain that occurs during the process. An annealing under vacuum (650°C, 6h) was performed to further reduce the residual twist angle [2] and increase the domain size. The KFM image after the annealing is presented in Figure S1 (b). The linescan in Figure S1 (c) shows a contrast of about 200 mV between AB and BA domains.

The charge tunable device presented in Figure S1 (d) was fabricated with the method described in ref. [3]. The flakes are transferred on a SiO$_2$/Si substrate with Ti/Au electrodes pre-patterned by photolithography. Figure S1 (d) presents a sketch of the structure and indicates the flakes thicknesses. The hBN crystals comes from the reference synthesis method at High Pressure High Temperature [4]. The WSe$_2$ ML was exfoliated from a bulk crystal purchased at 2D Semiconductor. Two graphite flakes were exfoliated from a HOPG bulk crystal. The first one is the bottom gate and it is also in contact with a Ti/Au electrode. The second one is used to contact the WSe$_2$ ML to another Ti/Au electrode. The ferroelectric interface is localized between the two green flakes in Figure S1 (d), and is 6 nm away from the ML WSe$_2$. The KFM image measured on the full structure, which gives the surface potential above the top hBN (see Figure S1 (d)), is shown in Figure S1 (e), with a line scan along the red line in (f). The contrast of the KFM image is suited to distinguish clearly the AB domain (dark blue) and BA domain (light blue) that emerge from the FE interface.

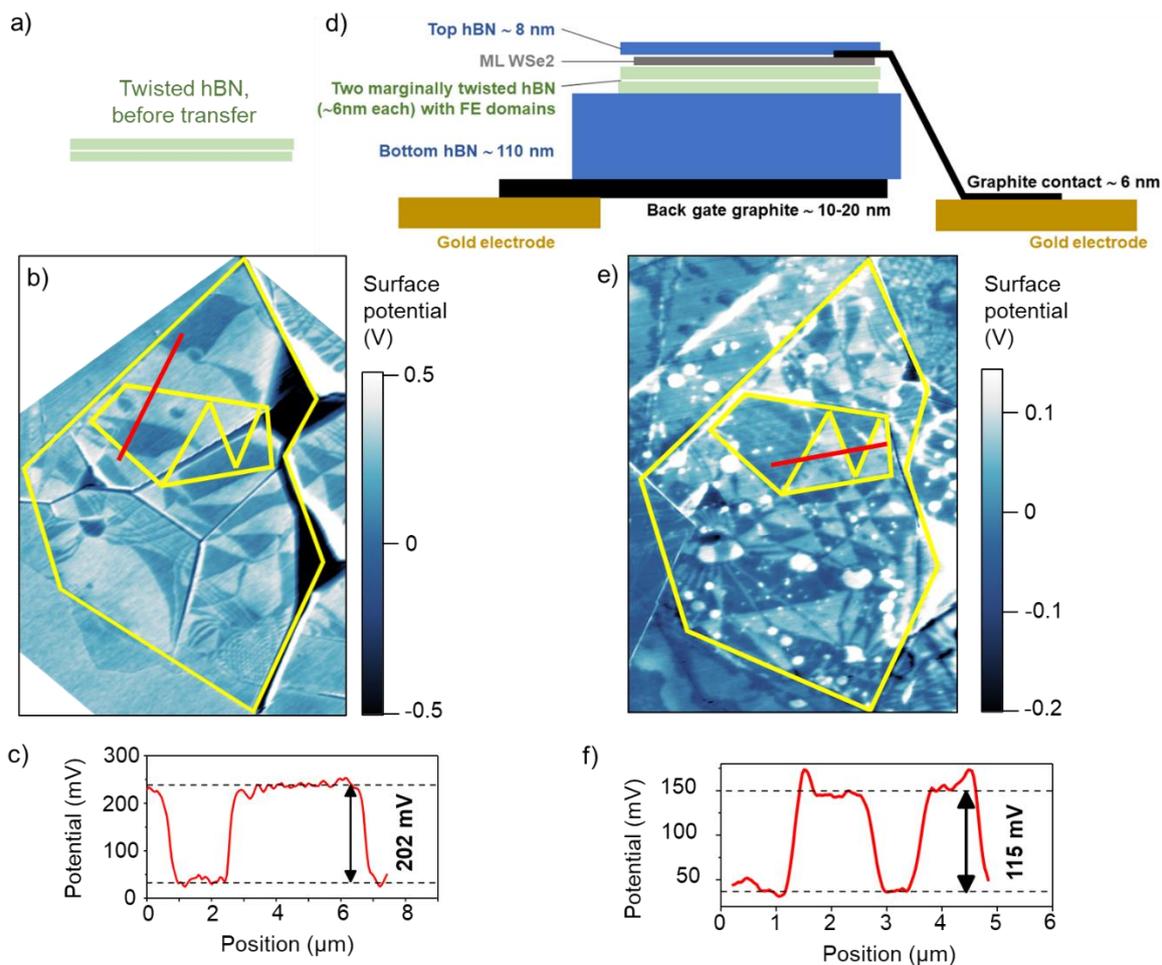

*Figure S1 (a) Sketch, (b) KFM image and (c) linescan of the surface potential of the FE hBN structure before it was transferred on the whole structure. The linescan was performed along the red line in (b). (d) Sketch of the whole structure, (e) KFM image made on the top hBN, (f) linescan of the surface potential along the red line in (e). The position of the bigger FE domain after the transfer are depicted by solid yellow lines in (b) and (e).*

The comparison of the KFM images before and after the transfer (Figure S1 (b) and (e)) reveals several features. First, we observe that the shape of the domains changes during the transfer, likely due to strain induced by the transfer process. Second, the contrast in surface potential between the AB and BA domains is reduced by almost half following the transfer. We attribute this decrease to the screening of the potential of the buried FE interface, which is covered by both the $WSe_2$ monolayer and the top hBN layer. Lastly, the full structure image clearly shows saturation with white spots of high surface potential, which we believe are caused by polymer residues from the flake transfer and by bubbles trapped between the layers.

Fig. S2 shows optical images of the full stack.

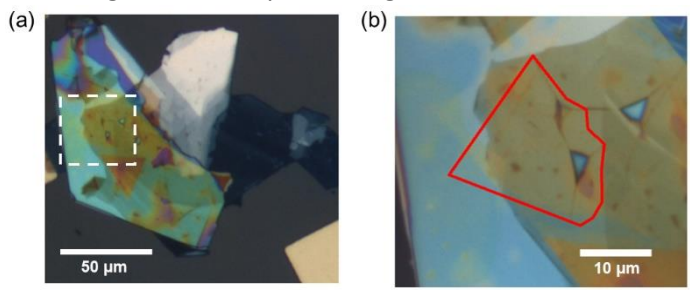

*Figure S2 -Optical image of the structure with x4 objective (a) and with a x20 objective (b). The position of image with x20 objective is marked in (a) with dashed line. The position of the $WSe_2$ ML is indicated with a solid red line in (b).*

**Details on the experiments :**

The ferroelectricity of the hBN interface was probed with Kelvin probe Force Microscopy (KFM) at 300K with CSI instrument (Nano-Observer microscope and an AFM tip made of Si and coated with Pt) as detailed in ref [5]. The KFM measurements shown in the main manuscript were performed on the whole structure.

The micro-photoluminescence experiments were performed with a continuous wave HeNe laser excitation in an ultra-stable confocal microscope described in ref. [5]. For the mapping, the sample is moved with respect to the laser spot using a xyz piezodriven scanner from Attocube. Together with the high numerical aperture of the objective (NA=0.82), it guarantees a spatial resolution of about 500 nm.

**Carrier density as a function of the gate voltage**

The carrier density is estimated from the simple plate capacitance model. A change of bias voltage $\Delta V$ induce a change of the electron density $\Delta n$ by:

$$\Delta n = \frac{\varepsilon_0 \varepsilon_{hBN}}{e * t} \Delta V$$

With $\varepsilon_0$ vacuum permittivity, $\varepsilon_{hBN}$ the out-of-plane dielectric permittivity of hBN, measured at 3.4 [6], $t$ the thickness of hBN between the WSe$_2$ and the bottom gate (122 nm, considering the bottom hBN and the twisted hBN with FE domains, see Figure S1 (d)), and $e$ the elementary charge.

**Large range colormap to probe the high doping regime:**

While the optical detection of the hBN FE states with the WSe$_2$ ML is shown to be efficient close to neutrality (between +/- 0.5 V, see Fig. 2 in the main text), our setup also allows to probe the high doping regime by applying higher voltage between the two gates.

The Figure S3 shows color maps made with spectra acquired from +20V to -20V (left, BA domain impose at the beginning of the scan), and with spectra acquired from -20 to +20V (right) (AB domains imposed at the beginning of the scan). No clear distinction can be found between the two plots, showing that the effect of the FE hBN is negligible at high doping.

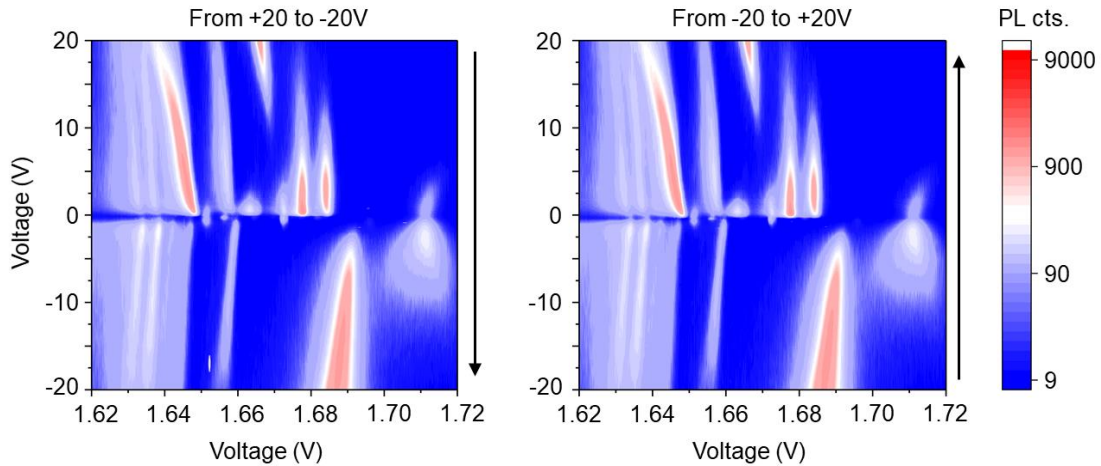

*Figure S3 - PL spectra as a function of gate voltage, measured at 1µW, 5K.*

**PL as function of the gate voltage on a sample without FE domains:**

Figure S4 displays photoluminescence (PL) data obtained from a reference sample, similar to the one analyzed in the main text, but without ferroelectric (FE) domains. Panels (a) and (b) show bias-dependent PL measurements in the voltage ranges of -15V to +15V and +15V to -15V, respectively. These are comparable to the plot shown in Figure S3. Close-up views near the neutrality point are provided in panels (c) and (d). Panel (e) plots the ratio of $X_S^-/X^0$, revealing no noticeable difference between the measurements taken with a +15V or -15V gate voltage applied before the measurement. Finally, two spectra at 0V are presented in panel (f), one measured after applying +15V and the other after applying -15V. Figures S4 (e-f) demonstrate that no hysteretic behavior is observed in the absence of FE domains.

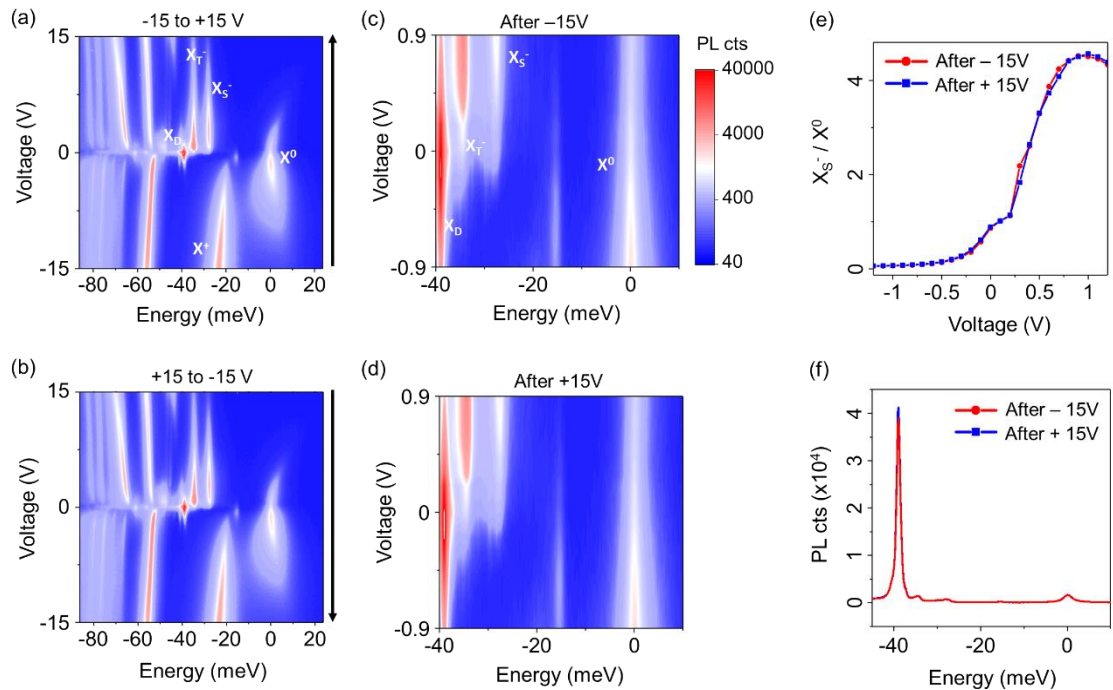

*Figure S4: PL spectra as a function of gate voltage from -15 to +15V in (a) and from -15 to +15V in (b). (c-d) zoom of (a) in (c), and zoom of (b) in (d). (e) PL intensity ratio $X_S^-/X^0$ as function of the gate voltage*

Hexagonal Boron Nitride. *Mater. Res. Express* **2022**, *9* (6), 065901. https://doi.org/10.1088/2053-1591/ac4fe1.